\documentclass[aps,twocolumn,prl,showkeys,preprintnumbers,superscriptaddress,nobibnotes,floatfix,notitlepage,nofootinbib]{revtex4-2}

\usepackage{graphicx}
\usepackage{dcolumn}
\usepackage{bm}
\usepackage[dvipsnames]{xcolor}
\usepackage{enumitem}

\begin{document}

\preprint{LA-UR-25-26467}

\title{First Event-by-Event Identification of Cherenkov Radiation from Sub-MeV Particles in Liquid Argon}
\affiliation{Bartoszek~Engineering,~Aurora,~IL~60506,~USA}
\affiliation{Columbia~University,~New~York,~NY~10027,~USA}
\affiliation{University~of~Edinburgh,~Edinburgh,~United~Kingdom}
\affiliation{Embry$-$Riddle~Aeronautical~University,~Prescott,~AZ~86301,~USA }
\affiliation{University~of~Florida,~Gainesville,~FL~32611,~USA}
\affiliation{Los~Alamos~National~Laboratory,~Los~Alamos,~NM~87545,~USA}
\affiliation{Massachusetts~Institute~of~Technology,~Cambridge,~MA~02139,~USA}
\affiliation{Universidad~Nacional~Aut\'{o}noma~de~M\'{e}xico,~CDMX~04510,~M\'{e}xico}
\affiliation{University~of~New~Mexico,~Albuquerque,~NM~87131,~USA}
\affiliation{New~Mexico~State~University,~Las~Cruces,~NM~88003,~USA}
\affiliation{Texas~A$\&$M~University,~College~Station,~TX~77843,~USA}

\author{A.A.~Aguilar-Arevalo}
\affiliation{Universidad~Nacional~Aut\'{o}noma~de~M\'{e}xico,~CDMX~04510,~M\'{e}xico}
\author{S.~Biedron}
\affiliation{Element~Aero,~San~Leandro,~CA~94577,~USA}
\author{J.~Boissevain}
\affiliation{Bartoszek~Engineering,~Aurora,~IL~60506,~USA}
\author{M.~Borrego}
\affiliation{Los~Alamos~National~Laboratory,~Los~Alamos,~NM~87545,~USA}
\author{L.~Bugel\textsuperscript{\dag}}
\affiliation{Massachusetts~Institute~of~Technology,~Cambridge,~MA~02139,~USA}
\author{M.~Chavez-Estrada}
\affiliation{Universidad~Nacional~Aut\'{o}noma~de~M\'{e}xico,~CDMX~04510,~M\'{e}xico}
\author{J.M.~Conrad}
\affiliation{Massachusetts~Institute~of~Technology,~Cambridge,~MA~02139,~USA}
\author{R.L.~Cooper}
\affiliation{Los~Alamos~National~Laboratory,~Los~Alamos,~NM~87545,~USA}
\affiliation{New~Mexico~State~University,~Las~Cruces,~NM~88003,~USA}
\author{J.R.~Distel}
\affiliation{Los~Alamos~National~Laboratory,~Los~Alamos,~NM~87545,~USA}
\author{J.C.~D’Olivo}
\affiliation{Universidad~Nacional~Aut\'{o}noma~de~M\'{e}xico,~CDMX~04510,~M\'{e}xico}
\author{E.~Dunton}
\affiliation{Los~Alamos~National~Laboratory,~Los~Alamos,~NM~87545,~USA}
\author{B.~Dutta}
\affiliation{Texas~A$\&$M~University,~College~Station,~TX~77843,~USA}
\author{D.E.~Fields}
\affiliation{University~of~New~Mexico,~Albuquerque,~NM~87131,~USA}
\author{M.~Gold}
\affiliation{University~of~New~Mexico,~Albuquerque,~NM~87131,~USA}
\author{E.~Guardincerri}
\affiliation{Los~Alamos~National~Laboratory,~Los~Alamos,~NM~87545,~USA}
\author{E.C.~Huang}
\affiliation{Los~Alamos~National~Laboratory,~Los~Alamos,~NM~87545,~USA}
\author{N.~Kamp}
\affiliation{Massachusetts~Institute~of~Technology,~Cambridge,~MA~02139,~USA}
\author{D.~Kim}
\affiliation{University of South Dakota, Vermillion, SD 57069, USA}
\author{K.~Knickerbocker}
\affiliation{Los~Alamos~National~Laboratory,~Los~Alamos,~NM~87545,~USA}
\author{W.C.~Louis}
\affiliation{Los~Alamos~National~Laboratory,~Los~Alamos,~NM~87545,~USA}
\author{C.F.~Macias-Acevedo}
\affiliation{Universidad~Nacional~Aut\'{o}noma~de~M\'{e}xico,~CDMX~04510,~M\'{e}xico}
\author{R.~Mahapatra}
\affiliation{Texas~A$\&$M~University,~College~Station,~TX~77843,~USA}
\author{J.~Mezzetti}
\affiliation{University~of~Florida,~Gainesville,~FL~32611,~USA}
\author{J.~Mirabal}
\affiliation{Los~Alamos~National~Laboratory,~Los~Alamos,~NM~87545,~USA}
\author{M.J.~Mocko}
\affiliation{Los~Alamos~National~Laboratory,~Los~Alamos,~NM~87545,~USA}
\author{D.A.~Newmark}\email{Contact author: dnewmark@mit.edu}
\affiliation{Massachusetts~Institute~of~Technology,~Cambridge,~MA~02139,~USA}
\author{P.~deNiverville}
\affiliation{Los~Alamos~National~Laboratory,~Los~Alamos,~NM~87545,~USA}
\author{C.~O’Connor}
\affiliation{University~of~Florida,~Gainesville,~FL~32611,~USA}
\author{V.~Pandey}
\affiliation{Fermi National Accelerator Laboratory, Batavia, Illinois 60510, USA}
\author{D.~Poulson}
\affiliation{Los~Alamos~National~Laboratory,~Los~Alamos,~NM~87545,~USA}
\author{H.~Ray}
\affiliation{University~of~Florida,~Gainesville,~FL~32611,~USA}
\author{E.~Renner}
\affiliation{Los~Alamos~National~Laboratory,~Los~Alamos,~NM~87545,~USA}
\author{T.J.~Schaub}
\affiliation{Los~Alamos~National~Laboratory,~Los~Alamos,~NM~87545,~USA}
\author{A.~Schneider}
\affiliation{Los~Alamos~National~Laboratory,~Los~Alamos,~NM~87545,~USA}
\author{M.H.~Shaevitz}
\affiliation{Columbia~University,~New~York,~NY~10027,~USA}
\author{D.~Smith}
\affiliation{Embry$-$Riddle~Aeronautical~University,~Prescott,~AZ~86301,~USA }
\author{W.~Sondheim}
\affiliation{Los~Alamos~National~Laboratory,~Los~Alamos,~NM~87545,~USA}
\author{A.M.~Szelc}
\affiliation{University~of~Edinburgh,~Edinburgh,~United~Kingdom}
\author{C.~Taylor}
\affiliation{Los~Alamos~National~Laboratory,~Los~Alamos,~NM~87545,~USA}
\author{A.~Thompson}
\affiliation{Northwestern~University,~Evanston,~IL~60208,~USA}
\author{W.H.~Thompson}
\affiliation{Los~Alamos~National~Laboratory,~Los~Alamos,~NM~87545,~USA}
\author{M.~Tripathi}
\affiliation{University~of~Florida,~Gainesville,~FL~32611,~USA}
\author{R.T.~Thornton}
\affiliation{Los~Alamos~National~Laboratory,~Los~Alamos,~NM~87545,~USA}
\author{R.~Van~Berg}
\affiliation{Bartoszek~Engineering,~Aurora,~IL~60506,~USA}
\author{R.G.~Van~de~Water}
\affiliation{Los~Alamos~National~Laboratory,~Los~Alamos,~NM~87545,~USA}

\begingroup
\renewcommand\thefootnote{\dag}
\footnotetext{Deceased}
\endgroup

\collaboration{The CCM Collaboration}

\date{\today}

\begin{abstract}

This Letter reports the event-by-event observation of Cherenkov light from sub-MeV electrons in a high scintillation light-yield liquid argon (LAr) detector by the Coherent CAPTAIN-Mills (CCM) experiment. The CCM200 detector, located at Los Alamos National Laboratory, instruments 7~tons (fiducial volume) of LAr with 200 8-inch photomultiplier tubes (PMTs), 80\% of which are coated in a wavelength shifting material and the remaining 20\% are uncoated. In the prompt time region of an event, defined as $-6 \leq t < 0$~ns relative to the event start time $t=0$, the uncoated PMTs are primarily sensitive to visible Cherenkov photons. Using gamma-rays from a $^{22}$Na source for production of sub-MeV electrons, we isolated prompt Cherenkov light with $>5\sigma$ confidence and developed a selection to obtain a low-background electromagnetic sample. This is the first event-by-event observation of Cherenkov photons from sub-MeV electrons in a high-yield scintillator detector, and represents a milestone in low-energy particle detector development.

\end{abstract}

\maketitle

\paragraph{\label{sec:intro}Introduction---}
The study of weakly interacting physics and rare event searches requires large and ultra-sensitive detectors to separate signals from backgrounds. \textit{Optical detectors} consisting of bulk material transparent to visible light and instrumented with photomultiplier tubes (PMTs) are a cost-effective way to create ton to kiloton-scale experiments for observing low-energy or rare processes. These detectors reconstruct events from the scintillation light or Cherenkov radiation in the medium. Detectors utilizing a high light-yield scintillator, which typically produces $\mathcal{O}(10^4)$ visible photons per MeV of deposited energy, have excellent energy resolution even for low particle kinetic energies, but are unable to reconstruct directional information with only the isotropic scintillation photon emission~\cite{BOREXINO:2020aww,KamLAND:2013rgu,JUNO:2024jaw}. Detectors designed for reconstruction of Cherenkov radiation, on the other hand, which is emitted at a characteristic angle with respect to the tracks of relativistic particles, are able to reconstruct directional information and discriminate between particle types using Cherenkov ring topology~\cite{Patterson:2009ki,Super-Kamiokande:2019gzr}. However, such detectors have poorer energy resolution and higher kinetic energy detection thresholds because the Cherenkov photon yield is $\mathcal{O}(10^2)$ smaller than that of high light-yield liquid scintillators~\cite{SNO:2024vjl,Super-Kamiokande:2024qbv,MiniBooNE:2020pnu}.

A high light-yield scintillation \textit{hybrid optical detector} with the ability to simultaneously reconstruct Cherenkov and scintillation signals has been a long-standing goal for MeV-scale particle physics instrumentation~\cite{Aberle:2013jba}. Community studies such as the 2021 Snowmass Process highlight the importance of hybrid systems with event-by-event Cherenkov and scintillation separation for rare event searches and high-priority particle physics goals~\cite{Escobar:2022jau,Klein:2022lrf}. In neutrinoless double beta decay, Cherenkov light enables background rejection via back-to-back electron identification, while precise energy resolution resolves the decay endpoint. For solar neutrinos, directional reconstruction using Cherenkov light helps identify electrons traveling in the direction of the sun, and low energy thresholds are needed to detect sub-MeV events. In recent years, we have seen important steps in demonstrating practical designs from two oil-based optical detectors for solar neutrino experiments. The Borexino collaboration has presented a statistical (as opposed to event-by-event) observation of Cherenkov light for sub-MeV solar neutrino events~\cite{PhysRevLett.128.091803}. SNO+ reported the first event-by-event direction reconstruction of $>5$ MeV solar neutrinos in a high-yield scintillator detector~\cite{PhysRevD.109.072002}.

In this Letter, we present the first event-by-event observation of Cherenkov radiation in a high light-yield scintillation detector from electrons with kinetic energy $\lesssim 1$~MeV. This work utilizes the CCM200 detector, a 10~ton liquid argon (LAr) light collection only detector with 50\% photocoverage and 2~ns timing resolution; and it specifically analyzes data from a $^{22}$Na calibration source, which emits gamma-ray photons that Compton scatter to create relativistic electrons. Because Cherenkov light is prompt, with an $\mathcal{O}(\rm{ps})$ production time constant, it produces PMT signals before the reconstructed event start time at $t=0$. We present a detailed model of the time dependence of the Cherenkov signal, purity of Cherenkov radiation in the selected time region, and the angle between pairs of signals in this Cherenkov enhanced region.

The use of LAr, a high-yield scintillator capable of producing an average of 40,000 photons per MeV of deposited energy, offers a novel approach for hybrid MeV-scale detectors~\cite{Doke:1990rza}. Scintillation light is produced by the decay of excited argon dimers with singlet and triplet spin configurations. These states have two characteristic decay time constants, $\mathcal{O}(5{\rm ~ns})$ and $\mathcal{O}(1000{\rm ~ns})$, respectively~\cite{Whittington:2014aha,Segreto:2020qks,DEAP:2020hms}. For both states, the scintillation photon emission spectrum is narrowly peaked around 128~nm vacuum wavelength~\cite{Heindl:2010zz}. For lightly ionizing electrons, approximately 20\%-30\% of the light is emitted by the singlet state, the rest by the longer-lived triplet state. Thus, although purified LAr produces approximately four times more scintillation light than organic scintillators, reducing the total Cherenkov-to-scintillation photon ratio, the prompt Cherenkov light appearing in the first few nanoseconds of the event can be as well-separated as in fast oil-based scintillators given the proportion of short-lived singlet to long-lived triplet excimer states. Additionally, the use of LAr provides several major advantages described below.
\begin{enumerate}[align=left, label=\arabic*., labelwidth=0pt, labelsep=0.2em, leftmargin=0.0em, itemindent=0pt]
\item Unlike oil-based detectors, pure LAr does not intrinsically absorb optical photons with wavelengths greater than the first excimer continuum of $\lambda=113.2$~nm~\cite{Neumeier:2012cz}. Therefore, LAr detectors do not require bulk wavelength-shifting dopants. Instead, ultraviolet (UV) photons in pure LAr propagate without absorption and can be converted to visible light by wavelength-shifter coatings on edges of the detector. Thus, in contrast to oil, where short-wavelength photons are absorbed and isotropically re-emitted by dopants near the interaction site, UV Cherenkov light in LAr maintains its directionality as it travels to the detector walls. While this study focuses on detecting visible Cherenkov photons, future work could explore detection of UV Cherenkov photons using PMTs with picosecond-level timing for separation from scintillation photons. Given that Cherenkov emission scales as $1/\lambda^2$, capturing shorter-wavelength photons could significantly enhance directional reconstruction, even for sub-MeV electrons. This capability would especially be valuable for applications such as low-energy solar neutrino detection.

\item Two factors delay scintillation light arrival at the PMTs in LAr compared to Cherenkov light, which will be key to their separation: (1) the intrinsic fast scintillation decay time constant of $\mathcal{O}(5~\mathrm{ns})$ is slower than typical scintillation oil time constants of $\mathcal{O}(3~\mathrm{ns})$~\cite{Elisei:1997tw,Borexino:2008gab}; and (2) wavelength shifting material instrumentation can introduce additional propagation delays. In this work, 20\% of PMTs are not coated in a wavelength shifter, requiring that UV scintillation light must propagate to a wavelength shifter on the walls of the detector or coated PMTs to wavelength shift then propagate back into an uncoated PMT for detection. This increases the time delay between prompt visible Cherenkov photons and delayed wavelength shifted scintillation photons in the uncoated PMTs, enabling this analysis.

\item Lastly, although the cryogenic nature of LAr detectors introduces complexity, it also offers an additional benefit of lower dark current rates, if, in the future, one uses Silicon photomultipliers (SiPMs). SiPMs are characterized by lower operating voltages, cost, and radioactive backgrounds compared to traditional PMTs~\cite{Wang:2022zsv}, and are an active area of development in particle physics readout technologies. One of the large obstacles in SiPM deployment for physics studies at room temperature is the dark current rate. At LAr cryogenic temperatures, however, the dark current rate drops by several orders of magnitude ~\cite{Wang:2022zsv,Anfimov:2020ikk,Ozaki:2020liq}. Lower dark rates, along with active research into broader wavelength sensitivity, would allow for cleaner separation of the Cherenkov signal from backgrounds using SiPMs in LAr--- an advantage not present in oil-based systems.
\end{enumerate}

While LAr has some disadvantages as a medium for hybrid Cherenkov and scintillation detectors--- it is more complex to handle than room-temperature scintillating oils, the overall lower index of refraction compared to oil reduces the Cherenkov light yield, and inherent scattering may impact directional reconstruction--- these can be outweighed by the previously listed advantages, along with the decades of research into LAr and the lower threshold for Cherenkov radiation of $\sim$0.2~MeV kinetic energy for electrons. In this study, we demonstrate that selecting the earliest signals on the uncoated PMTs allows for reliable tagging of Cherenkov light--- a signature of electromagnetic physics that can be leveraged for background rejection. This novel application of optical LAr detection is made possible by the advantages discussed above and improved understanding of the detector technology.

\paragraph{\label{sec:detector}The CCM200 Detector---}

The CCM200 detector was constructed at the Lujan Center at the Los Alamos Neutron Science Center (LANSCE) beamline to study MeV-scale neutrinos and beyond Standard Model (BSM) physics~\cite{Tripathi:2024jnq,Dunton:2022dez}. The two primary signal categories isolate 1) MeV-scale electromagnetic final states with no accompanying highly ionizing signal and 2) highly ionizing $\mathcal{O}(100)$ keV coherent scattering signals that have no accompanying electromagnetic signal. CCM has already published world leading limits on BSM models from CCM120, the prototype detector for CCM200~\cite{CCM:2021yzc,CCM:2021leg,CCM:2021jmk,CCM:2023itc}. To further improve physics sensitivities with CCM200, both categories of analysis will benefit by enhanced signal-to-background separation through event-by-event Cherenkov light identification.

CCM200 is ideally suited for event-by-event Cherenkov photon identification. Its cryostat is an upright cylinder, 2.58~m in diameter and 2.25~m in height, holding 10~tons of LAr, split into a 7~ton inner fiducial region and a 3~ton optically isolated veto region. The walls of the inner detector are lined with Mylar reflective foils evaporatively coated in tetraphenyl butadiene (TPB) to wavelength shift UV photons into the visible spectrum for detection. CCM200 is instrumented with 200 8-inch R5912-Y002 10 stage cryogenic Hamamatsu PMTs, which provide 50\% photocoverage of the fiducial volume; 160 of these PMTs are evaporatively coated in TPB, while 40 PMTs were left uncoated to enhance rapid detection of the prompt Cherenkov light from electromagnetic signals. Event timing is critical to this analysis, and CAEN V1730/V1730S 500 MHz digitizer boards are used to read out the signals from the PMTs, providing digitization in 2~ns time bins with 14-bit precision. 

\paragraph{\label{sec:data} Detector Response Model and Data Selection---}

We process the time series of digitized PMT voltage differences offline to reconstruct observed ``hits''; these are produced by Cherenkov and scintillation photons in the detector inducing photoelectrons (PEs) on the PMT photocathode and thermal electrons emitted from the photocathode, also known as ``dark noise". The time-structure of these pulses is also affected by well-known post-pulsing effects common across many types of PMTs~\cite{MiniBooNE:2006fhd,Kaptanoglu:2017jxo,DEAP:2017fgw,Abbasi_2010,Brigatti_2005}. The accompanying paper, Ref.~\cite{companion_paper}, describes the methods by which we have fit the digitized waveforms for PEs as well as simulated and calibrated the complex time-structure using the Monte-Carlo event generator~\texttt{Geant4}~\cite{GEANT4:2002zbu}. This represents the first published description of the timing of light in a LAr detector that includes Cherenkov radiation. 

This work uses data from a 3~$\mu$Ci $^{22}$Na calibration source encapsulated in approximately 1~mm of stainless steel. The source is inserted on a stainless steel bayonet through a central flange at the top of the detector and placed at the midline of the detector. The $^{22}$Na isotope has two main decay pathways. The primary decay channel, with around 90\% branching ratio, is $\beta^+$ decay, emitting a 0.546~MeV positron and 1.275~MeV gamma-ray from subsequent de-excitation of $^{22}\rm{Ne}^*$~\cite{BASUNIA201569}. This positron promptly annihilates within the steel capsule and is not observed, but the annihilation creates two back-to-back 0.511~MeV gamma-rays that exit the source and enter the detector. The other decay channel, with approximately 10\% branching ratio, is electron capture, for which the signal is only the 1.275~MeV gamma-ray from $^{22}\rm{Ne}^*$ de-excitation. Hence, for these $^{22}$Na decay events, the source is emitting up to three gamma-rays--- two 0.511~MeV photons from positron annihilation and a single 1.275~MeV photon from nuclear de-excitation.

\begin{figure}[h] 
    \centering
    \includegraphics[width=\linewidth]{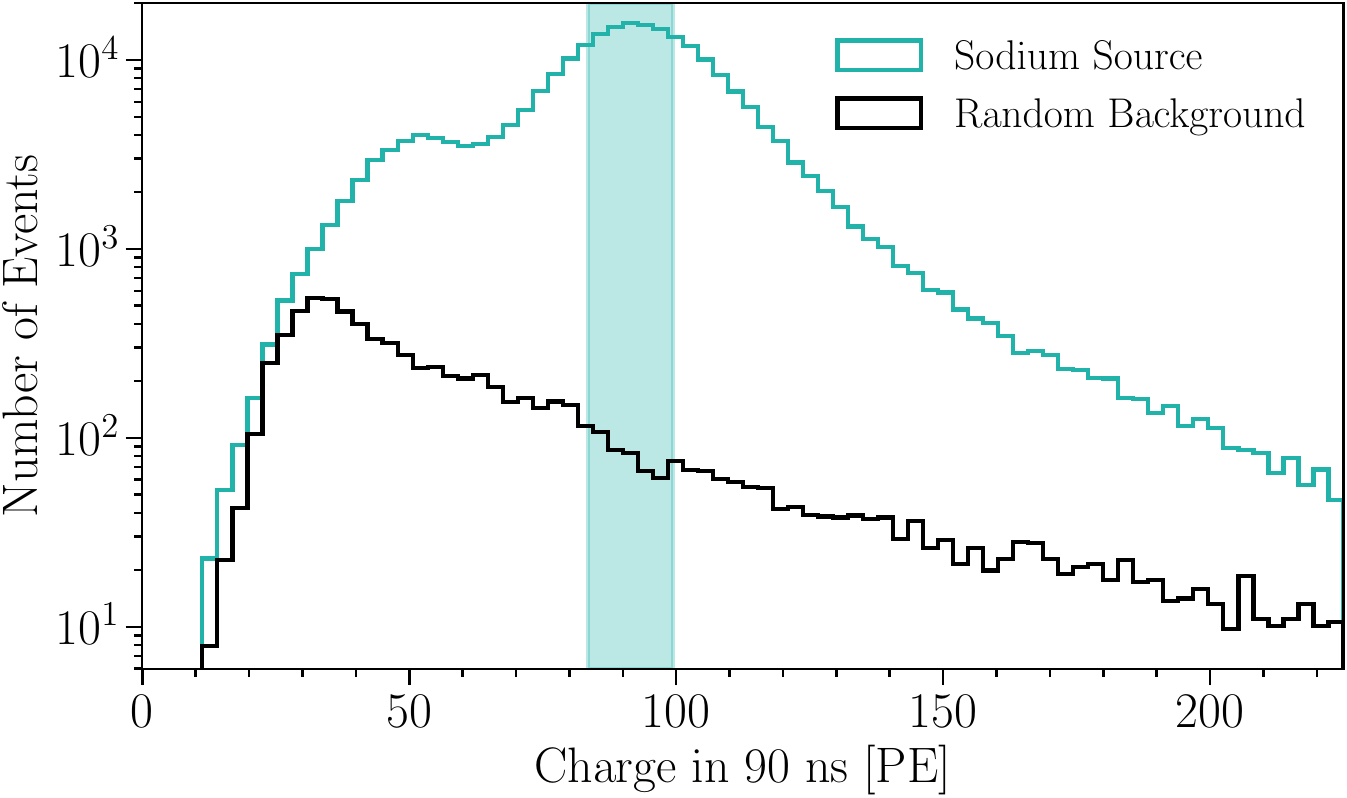} 
    \caption{Distribution of total charge in the first 90~ns of reconstructed events for data taken with the sodium source inserted at the origin of the detector (blue line) and data taken with the sodium source removed from the detector (black line). Shaded band shows the charge cut used to select sodium events for this analysis to reduce background contamination.}
    \label{fig:sodium_charge} 
\end{figure}

We collect data in a 16~$\mu$s data acquisition window using an external 20~Hz trigger. Within each trigger window, we reconstruct event start times by requiring the total charge in the detector to be over a certain threshold in a given time window. For this analysis, we used a charge threshold of 3~PE in a 2~ns time window to identify the event start time, $t = 0$. Because this study focuses on sub-MeV events with low Cherenkov photon emission, the early Cherenkov light generally does not pass the threshold. We apply data quality cuts that remove events with--- 1) high-charge cosmic muons observed within the 16~$\mu$s data acquisition window; 2) less than 2.2~$\mu$s separation between surrounding reconstructed events, since more than one event can occur in the trigger window; and 3) reconstructed radial positions greater than $25$~cm from the center of the detector (obtained by an average of PMT positions weighted by charge in the first 20~ns of an event)--- see Ref.~\cite{companion_paper} for details. Fig.~\ref{fig:sodium_charge} shows the distribution, after applying these cuts, of total charge in the first 90~ns of an event, for data with and without the $^{22}\rm{Na}$ source inserted. The two peaks due to the electron capture and $\beta^+$ decay pathways, at $\sim$50~PE and $\sim$100~PE respectively, can be identified. We select events with charges within $\pm8$~PE of the center of the high energy peak, corresponding to $\beta^+$ decay, to increase statistics and limit background contamination. With the data quality and charge cuts applied, we expect $\leq$ 0.57\% random background event contamination. 

\begin{figure}[h] 
    \centering
    \includegraphics[width=\linewidth]{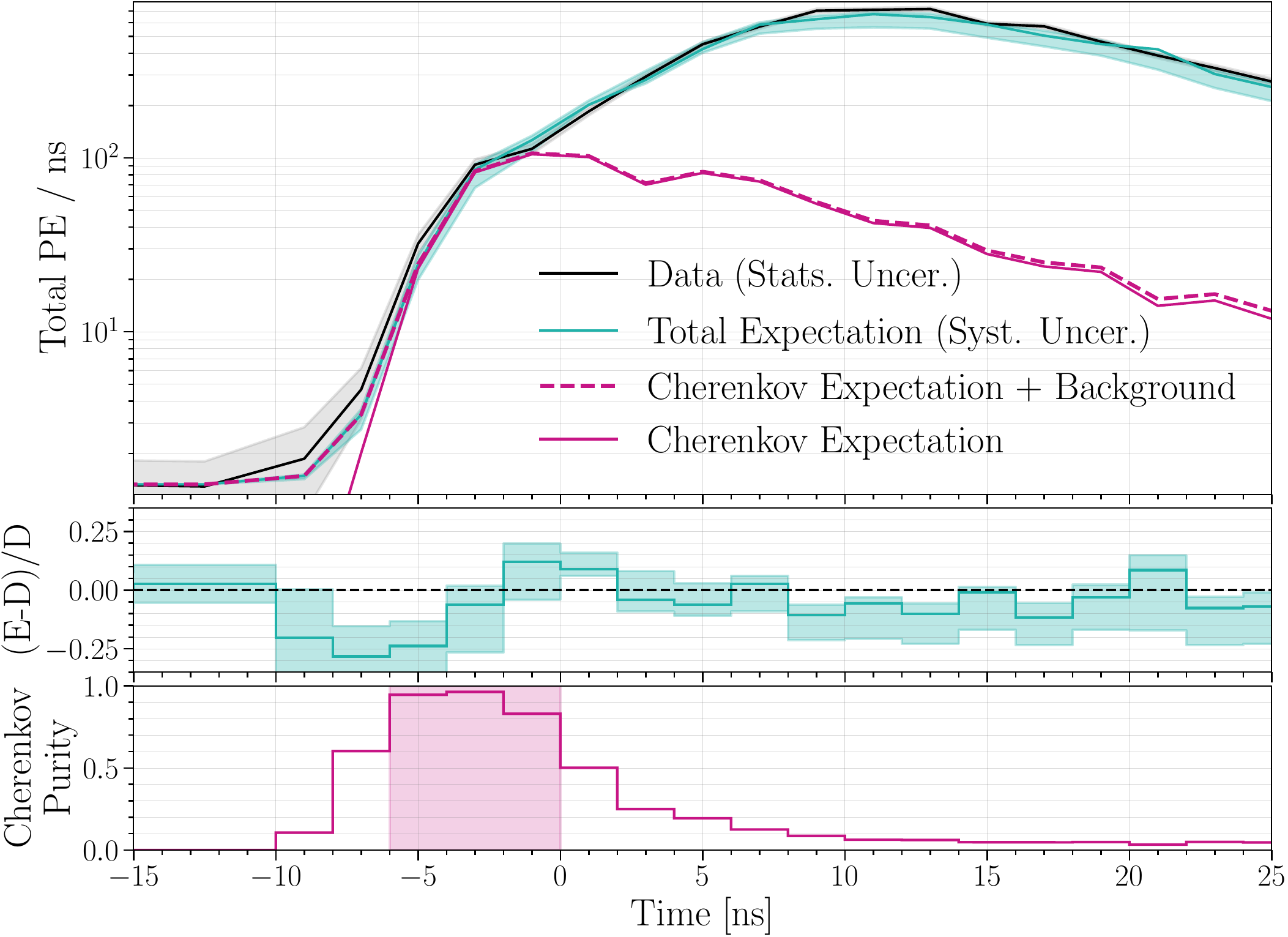} 
    \caption{The top plot shows the accumulated data (black line) from 43,522 selected sodium decay events for a typical uncoated PMT with a statistical uncertainty band as a function of time with respect to the reconstructed event start time. The corresponding simulation expectation is shown as a blue line with a systematic uncertainty band, which includes simulated scintillation and Cherenkov photons, and background photons measured when the source was removed. The magenta line is the Cherenkov photon contribution to the expectation. The middle plot shows the residual between simulation and data with uncertainties. The bottom plot shows ``Cherenkov purity", the ratio of expected Cherenkov photons to the total expectation.}
    \label{fig:typical_uncoated} 
\end{figure}

\paragraph{\label{sec:cherenkov}Cherenkov Separation---}
This analysis focuses on the early time region of the reconstructed PE pulse series from uncoated PMTs, since those are expected to have the clearest separation between prompt visible Cherenkov light and delayed wavelength shifted scintillation light. Fig.~\ref{fig:typical_uncoated}, top panel, shows the summed pulse series as a function of time $t$ of a typical uncoated PMT for 43,522 selected sodium decay events. The major feature that will be the primary focus of Cherenkov separation is clear in the data--- a small peak located at $t =-3$~ns, hence before the identified event start time. The behavior over the entire time region, which extends out to 2~$\mu$s after the start time, is discussed in Ref.~\cite{companion_paper}. The middle panel of Fig. 2 shows that within the 1$\sigma$ uncertainty band there is $\pm15\%$ or better agreement between simulation and data. The deviation between data and expectation centered around t = -7~ns is still under investigation.

\begin{figure}[h] 
    \centering
    \includegraphics[width=\linewidth]{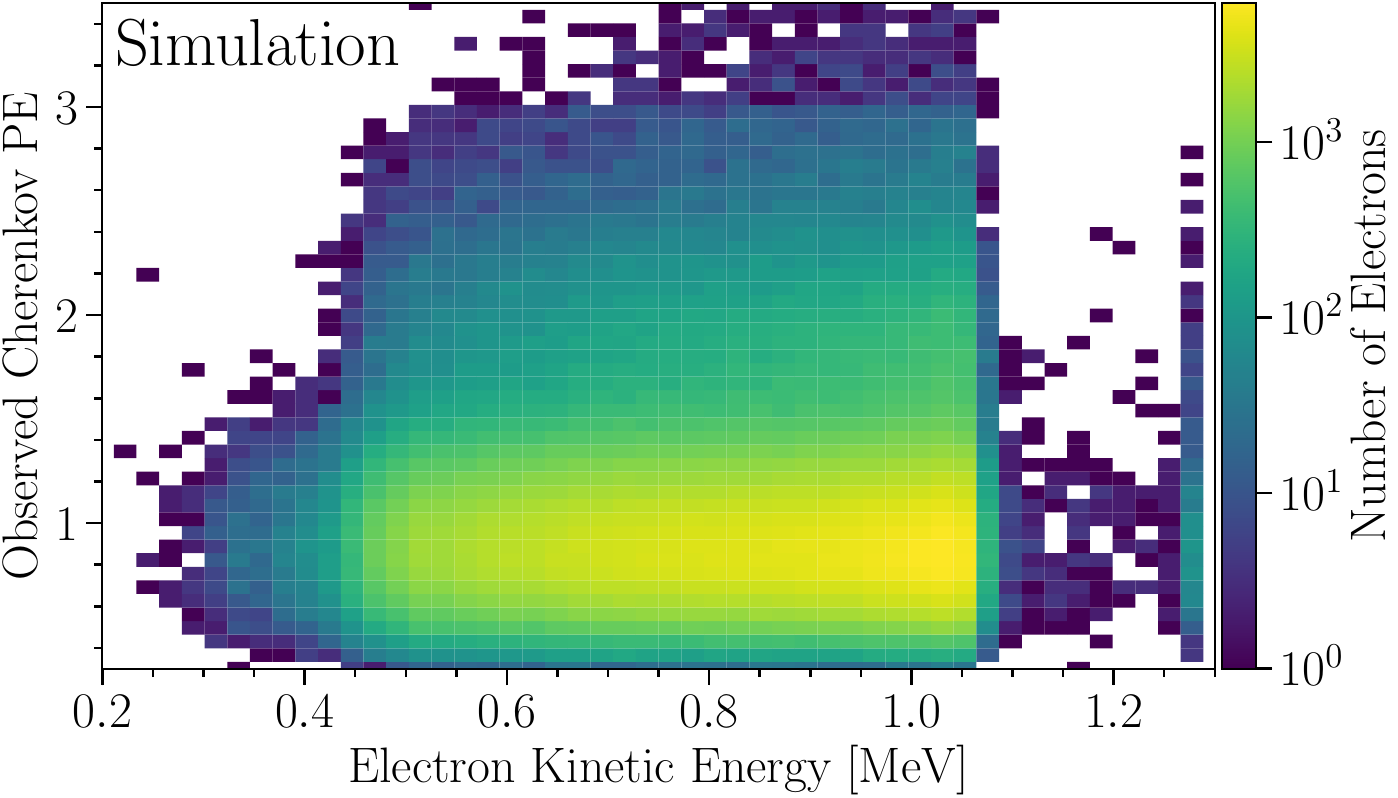} 
    \caption{Electron kinetic energy vs number of observed Cherenkov photons on the uncoated PMTs in the Cherenkov enhanced time region (color representing number of electrons). The majority of electrons that produce detectable Cherenkov photons have kinetic energies between $0.5 \lesssim KE \lesssim 1.0$~MeV, as expected for low energy photons that Compton scatter. Note the plot is zero suppressed.}
    \label{fig:energy_vs_cherenkov} 
\end{figure}

\begin{figure}[h] 
    \centering
    \includegraphics[width=\linewidth]{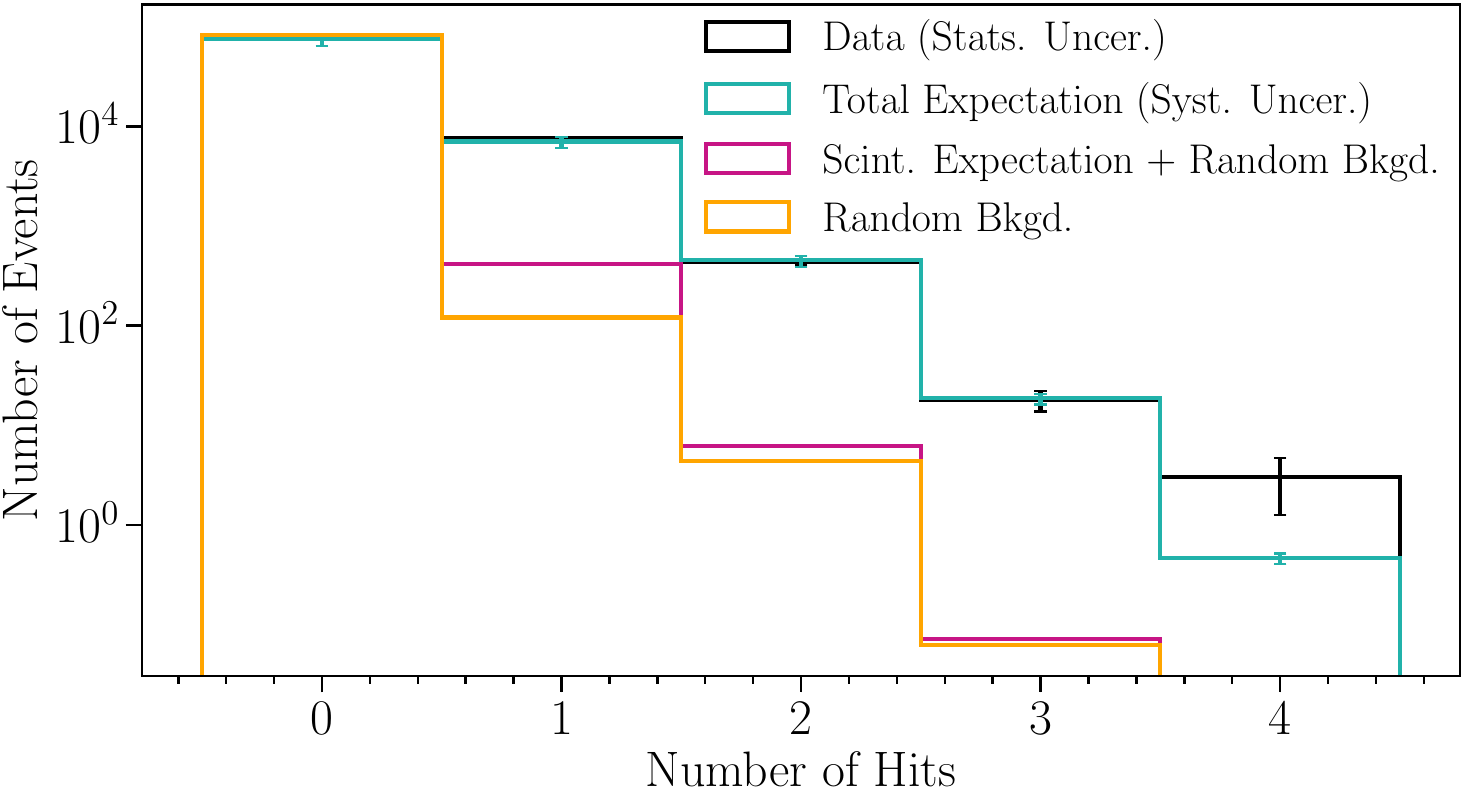} 
    \caption{Number of hits on the uncoated PMTs in the Cherenkov enhanced region. The data, with statistical uncertainties, is represented by the black line. The total simulation expectation, which combines simulated scintillation and Cherenkov photons with measured backgrounds, is represented by the blue line and has systematic uncertainties. The expected scintillation hits and random backgrounds only are the pink line. Finally, the random background only component of the expectation is the orange line.}
    \label{fig:nhits}
\end{figure}

\begin{figure}[h] 
    \centering
    \includegraphics[width=\linewidth]{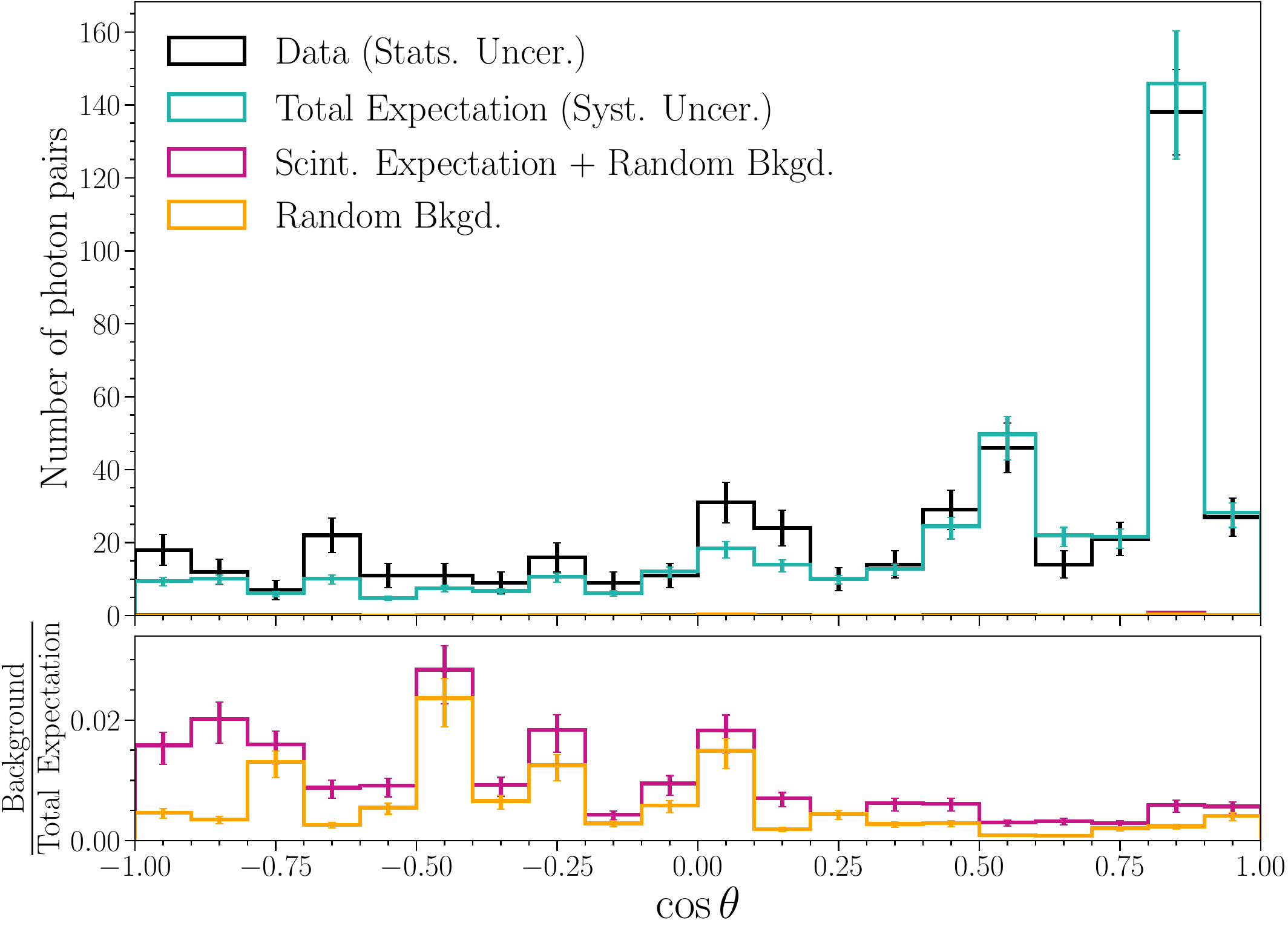} 
    \caption{Angle between photon pairs of hits in the early time region. Data (with statistical uncertainties) is represented by the black line, and the total expectation, combining Cherenkov, scintillation, and random background hits (with systematic uncertainties) is represented by the blue line. Data and simulation agree within $2\sigma$ uncertainties across all angles.}
    \label{fig:angular_data_vs_mc} 
\end{figure}

In Fig.~\ref{fig:typical_uncoated}, the small peak at $t=-3$~ns corresponds to a highly pure sample of Cherenkov photons, demonstrated by the ``Cherenkov purity'' shown in the lower plot. We will call this region of $-6 \leq t < 0$~ns the ``Cherenkov enhanced'' region and $t>0$ the ``scintillation enhanced'' region.  As expected, the scintillation enhanced region is delayed. The majority of photons originating from Cherenkov light appear before $t=0$; however, there is a substantial set of photons that are delayed because they are wavelength-shifted by the TPB and have longer paths to the PMTs. Fig.~\ref{fig:typical_uncoated} is the only direct fit between data and simulation, while the expectation in Figs.~\ref{fig:energy_vs_cherenkov},~\ref{fig:nhits}, and ~\ref{fig:angular_data_vs_mc} are the simulation output at the best fit parameters.

The 1.275~MeV and 0.511~MeV gamma-rays produced by the sodium source typically Compton scatter to produce electrons that cause scintillation, and will only emit Cherenkov light if above the $\sim$0.2~MeV kinetic energy threshold in LAr. Fig.~\ref{fig:energy_vs_cherenkov} is simulation and shows the correlation between electron kinetic energy and the observed number of Cherenkov photons on the uncoated PMTs in the Cherenkov enhanced time region, indicating that we can identify Cherenkov radiation produced from sub-MeV electrons. Note that the majority of electrons in the sodium decay events do not produce any observable Cherenkov PEs with this selection criteria, so the y-axis begins at 0.3~PE for legibility of the color scale. 

On an event-by-event basis, we can examine the hits in the Cherenkov enhanced region on the uncoated PMTs for the sodium decay dataset. Fig.~\ref{fig:nhits} shows the distribution of the number of hits in the Cherenkov enhanced region per event for the data and simulation (see caption for more details). We find $94.8\%$ purity of Cherenkov light in the Cherenkov enhanced region if $\geq 1$ hit is required.

With all cuts applied, including $\geq 1$ hit in the Cherenkov enhanced region on the uncoated PMTs, we find 9.78\% and 9.10\% selection efficiency in data and simulation, respectively. Given that this analysis only uses the uncoated PMTs, which provide 10\% total photocathode coverage in the detector, it is promising that this simple requirement of $\geq$ 1 hit in the Cherenkov enhanced region provides a high purity Cherenkov sample with adequate efficiency for identifying sub-MeV electromagnetic events. In the future, CCM can expand beyond this simple analysis to include information in the early part of the scintillation enhanced region and waveforms from coated PMTs, expecting to further improve the efficiency of isolating Cherenkov light in the early time region.

If there are at least two hits in the Cherenkov enhanced time region, we calculate the angle between the hits from the center of the hit PMTs and the source location (origin of the detector). This exploits the directionality of Cherenkov light compared to the isotropic scintillation and random backgrounds. Fig.~\ref{fig:angular_data_vs_mc} shows the calculated angles in the data and simulation. The data and total expectation have a clear preference for $0.8 < \cos\theta < 0.9$ as expected for visible Cherenkov photons produced by electrons with kinetic energies $0.7 \lesssim KE \lesssim 1.0$~MeV given the index of refraction in LAr is approximately $1.22$ for visible wavelengths~\cite{Sinnock:1969zz}. The scintillation and random background expectations are two orders of magnitude lower in rate and show a flatter angular distribution, as demonstrated by the residual of the ratio of the background-only expectations to the total expectation. The $\chi^2$ between data and the total expectation is 30.12 while the $\chi^2$ between data and the scintillation and random background only hypothesis is 473.60, both for 20 degrees of freedom. We reject the scintillation and random background only hypothesis using a $\Delta\chi^2$ test with $>5\sigma$ confidence, indicating event-by-event observation of Cherenkov light from sub-MeV electrons in LAr.

\paragraph{\label{sec:data_driven_validation}Data Driven Validation---}
Applying this analysis procedure to a data sample without any expected Cherenkov light further supports these findings. $^{57}$Co decays via electron capture and typically emits $122.06~\rm{keV}$ and $14.41~\rm{keV}$ gamma-rays from $^{57}$Fe de-excitation~\cite{Auble:1977pdy}. Both will scatter through the Compton and Photoelectric effects to produce electrons well below the Cherenkov radiation threshold. Using cobalt decay data collected during the same run period as the sodium calibration source, we have applied the same technique of selecting the Cherenkov enhanced time region and examining number of hits in the uncoated PMTs.

\begin{figure}[h] 
    \centering
    \includegraphics[width=\linewidth]{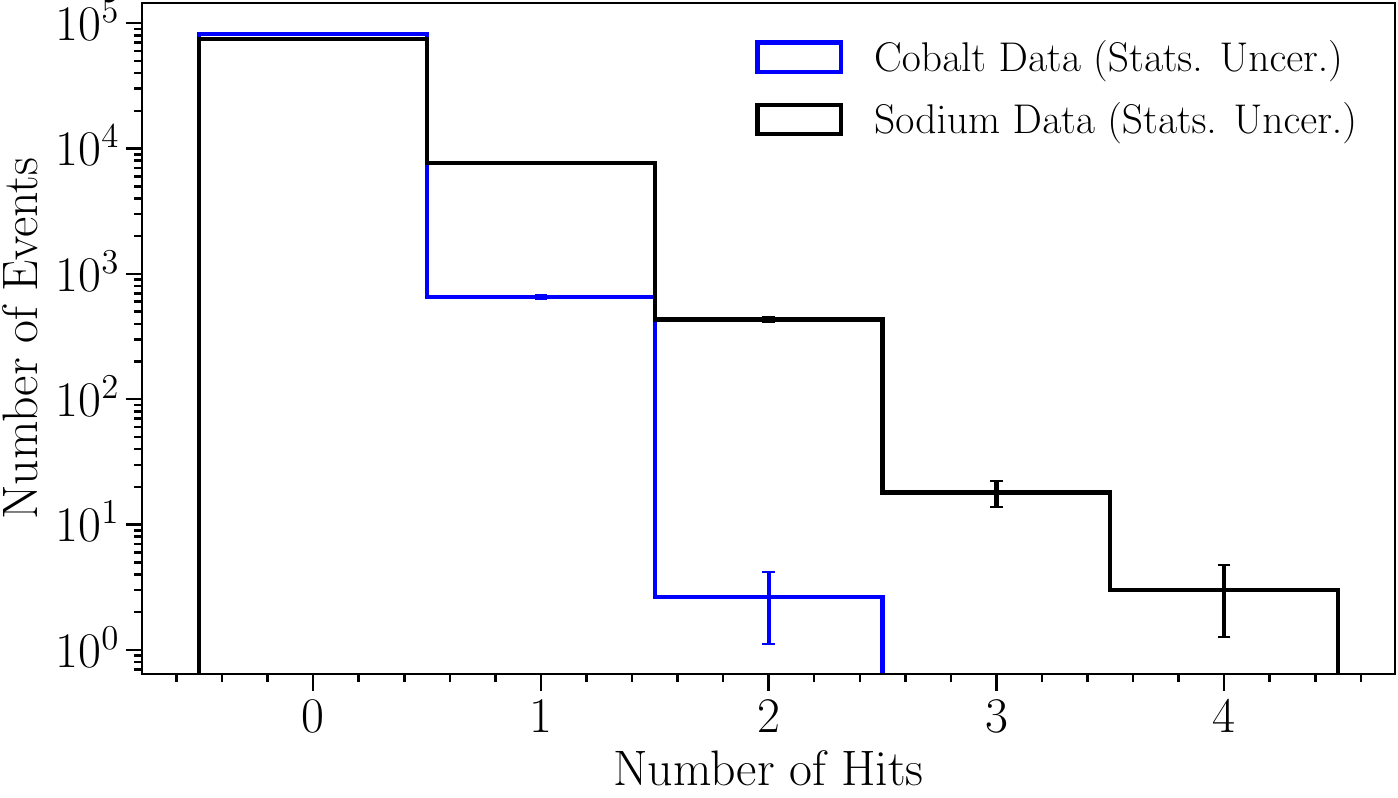} 
    \caption{Number of hits on the uncoated PMTs in the Cherenkov enhanced region. In the sodium data, Cherenkov light is expected to contribute hits in this early time region. For the cobalt data, which produces gamma-rays below Cherenkov threshold, hits in the early time region are only from scintillation photons or random backgrounds.}
    \label{fig:cobalt_nhits}
\end{figure}

Fig.~\ref{fig:cobalt_nhits} demonstrates the difference in rates of hits in the early time region on the uncoated PMTs in data with expected Cherenkov hits, sodium decays, and without, cobalt decays. For the sodium data, 9.78\% of events have $\ge~1$ hit while in the cobalt data, only 0.79\% of events have $\ge~1$ hit. This rate of hits in the cobalt data is in line with the expected scintillation and random background contamination demonstrated in Fig.~\ref{fig:nhits}, 0.51\% of events have $\ge~1$ hit.

\paragraph{\label{sec:conclusion}Conclusion---}
We have demonstrated observation of Cherenkov radiation on an event-by-event basis produced from sub-MeV electrons in a high-yield scintillator detector. This analysis was performed with CCM200, a LAr optical detector, which demonstrates a novel approach to building a hybrid detector that can separately reconstruct Cherenkov and scintillation signals. While experimental efforts are underway to engineer the ideal water or oil based liquid scintillator for Cherenkov separation, we have demonstrated the unique capabilities of LAr for this task~\cite{Anderson:2022lbb,Callaghan:2023oyu,Klein:2022tqr}. LAr is an ideal medium in many respects for hybrid detection--- pure LAr does not absorb optical photons, the scintillation emission time constants are relatively slow, and cryogenic conditions reduce the dark rate in certain detection technologies. The challenging aspects of separating low yield but broad spectrum Cherenkov light from high-yield UV scintillation photons can be overcome by a combining fast timing response and wavelength dependent detection mechanisms. Further studies investigating fast timing photo-detection and different methods of wavelength shifting light could further improve the ability to separate Cherenkov light from scintillation signals--- perhaps making LAr the ideal candidate for an ultra-large low-energy neutrino physics detector. 

\begin{acknowledgments}
We acknowledge the support of the Los Alamos National Laboratory LDRD and the U.S. Department of Energy Office of Science funding. We also wish to acknowledge support from the LANSCE Lujan Center and LANL’s Accelerator Operations and Technology (AOT) division. This research used resources provided by the Los Alamos National Laboratory Institutional Computing Program, which is supported by the U.S. Department of Energy National Nuclear Security Administration under Contract No.~89233218CNA000001. DAN is supported by the NSF Graduate Research Fellowship under Grant No.~2141064. AAA-A, CFMA, JCD, and MCE acknowledge support from DGAPA-UNAM Grant No.~PAPIIT-IN104723. We thank Josh Klein, University of Pennsylvania, for an interesting discussion on the benefits and drawbacks of LAr as a medium for a hybrid Cherenkov and scintillation detector.
\end{acknowledgments}

\bibliographystyle{apsrev4-2}
\bibliography{main}

\end{document}